\documentclass[a4paper]{cas-dc}

\usepackage[numbers]{natbib}
\usepackage{color}
\usepackage{tabularx}
\usepackage{booktabs}
\usepackage[frozencache, cachedir=./_minted]{minted}
\usepackage[most]{tcolorbox}
\usepackage{subcaption}
\usepackage{multirow}
\usepackage{pgfplots}
\pgfplotsset{compat=1.18}
\usepackage{circledsteps}
\usepackage{url}
\usepackage{xcolor}

\def\BibTeX{{\rm B\kern-.05em{\sc i\kern-.025em b}\kern-.08em
    T\kern-.1667em\lower.7ex\hbox{E}\kern-.125emX}}

\setminted{
    frame=lines,
    fontsize=\footnotesize,
    breaklines=true,
    autogobble,
    numbersep=2pt
}
\newtcolorbox{promptbox}[1]{
    title={\textsc{#1}},
    colframe=black!85,
    colback=black!2,
    coltitle=white,
    left=0.7em,
    right=0.7em,
    top=0.7em,
    bottom=0.7em,
    titlerule=0.5mm
}

\definecolor{rqbg}{HTML}{F3F3F3}      
\definecolor{rqtitle}{HTML}{444444}  
\newtcolorbox{rqsummary}[2][]{%
    enhanced,
    breakable,
    colback=rqbg,
    colframe=white,
    boxrule=0pt,
    arc=0pt,
    left=6mm,
    right=2mm,
    top=2mm,
    bottom=2mm,
    before skip=10pt,
    after skip=10pt,
    overlay={
    \fill[rqtitle] (frame.south west) rectangle ([xshift=5mm]frame.north west);
    \node[rotate=90, anchor=center, text=white, font=\sffamily\small] at ([xshift=2.5mm]frame.west) {#2};
    },
    {#1}
}

\newcommand{\bcircle}[1]{\Circled[fill color=black, inner color=white]{#1}}
\newcommand{\noindentparagraph}[1]{{\setlength{\parindent}{0pt} \setlength{\parskip}{0.5em} \textit{\textbf{#1}}\enspace}}

\begin{document}

\shorttitle{Legacy System Modernization with Coding Agents: A Case Study}
\shortauthors{Alves et al.}

\title[mode = title]{Legacy System Modernization with Coding Agents: A Case Study}

\author[1,3]{Iago da Silva Rodrigues Alves}
\ead{iagoalves@dcc.ufmg.br}

\author[2]{Cristiano Politowski}[orcid=0000-0002-0206-1056]
\ead{cristiano.politowski@ontariotechu.ca}

\author[3]{João Eduardo Montandon}[orcid=0000-0002-3371-7353]
\ead{joao@dcc.ufmg.br}

\cormark[1]
\fnmark[1]

\affiliation[1]{organization={Group Software},
    city={Belo Horizonte},
    state={Minas Gerais},
    country={Brazil}}

\affiliation[2]{organization={Ontario Tech University},
    city={Oshawa},
    state={Ontario},
    country={Canada}}

\affiliation[3]{organization={Universidade Federal de Minas Gerais (UFMG)},
    city={Belo Horizonte},
    state={Minas Gerais},
    country={Brazil}}

\cortext[1]{Corresponding author}

\begin{abstract}
    Legacy systems built on discontinued platforms are a recurring technological liability in organizations that depend on these applications to sustain critical business processes.
    Although modernization is strategically necessary, it is costly and error-prone when performed exclusively through manual effort.
    In this paper, we report on a case study conducted in a real industrial setting, where we evaluate both the effectiveness and efficiency of AI coding agents in supporting the migration of legacy systems to modern platforms.
    By using one version of Claude Code agent, we migrated 12 features of distinct complexity levels from a corporate ERP system written in Visual Basic 6 to C\# .NET 10 under a single-generation strategy.
    Once the migration sessions were completed, we measured the equivalence of the migrated features against the original ones, and collected data about the time spent and the number of tokens consumed by the agent during the process.
    In our experiment, the agent achieved an average equivalence of 70\%, with a strong asymmetry across complexity levels;
    low-level features achieved 92\%, and high-complexity ones scored 47\%.
    Similar asymmetry was observed in the cost-based metrics; low-level features consumed 1.47M tokens (\$1.66), whereas high-level ones used 9.09M (\$10.28).
    We reveal the practical scenarios and circumstances where this migration strategy is more effective, as well as discuss the limitations and challenges of using AI agents for legacy system modernization.
\end{abstract}


\maketitle

\section{Introduction}

Many organizations still run corporate systems built on languages and architectures that are now obsolete~\citep{Zaytsev2020}.
These legacy systems support critical business processes over time and became too big to be replaced or rewritten from scratch~\citep{Feathers2004}.
These organizations have no choice but to continuously maintain their codebases, posing a significant technological liability for their expansion~\citep{bisbal1999, Feathers2004}.

Maintaining legacy systems comes with several software development challenges, i.e.,~high maintenance costs, a reduced number of specialists in the original language, missing or outdated technical documentation, and incompatibility with modern components~\citep{Feathers2004,bisbal1999, Roziere2020, Mendonca2024}.
For example, Microsoft officially discontinued Visual Basic 6 (VB6) in 2008 and, since then, no longer provides security updates, making this environment increasingly risky~\citep{khadka2014icse}.
On the other hand, reengineering such projects takes too much time and money, and manually rewriting millions of lines of code can easily introduce behavioral differences from the original system~\citep{sneed1999wcre,Roziere2020, Mendonca2024}.

Large language models (LLMs) offer new possibilities here.
As they are extensively trained on large repositories of source code~\citep{Chen2021}, these models can understand syntax and semantics of programming languages, infer logical intent, and generate functional equivalents in other languages, making them useful tools for assisting or partially automating software migration~\citep{yang2024,hou2024, Macedo2025,Almeida2024, Alves2026,Islam2024}.
More recently, AI agents---which combine LLMs with tools and environment access to perform complex tasks autonomously---have emerged as a promising approach for software engineering tasks~\citep{Rondon2025, He2025a, Santos2026, Jimenez2024}.

\textbf{We report on a case study conducted in a real industrial setting}, where the first author was responsible for migrating two modules of a legacy corporate ERP system.
The system under migration---originally implemented in VB6---is in production for over two decades and currently supports more than 200 clients.
By using Claude Code agent under a single-generation strategy, we migrated 12 features of distinct complexity levels to C\# .NET 10, and compared the equivalence of legacy and migrated code, while also monitoring the agent's efficiency during the process.
Two research questions guided our investigation:

\begin{itemize}
    \item \textit{RQ1. To what extent is the migrated feature equivalent to the original one?}
          We measured equivalence in two dimensions: (a) \textit{persistence}, i.e.,~the migrated feature should persist the same data as the original one; and (b) \textit{functional behavior}, i.e.,~the migrated feature should implement the same business rules as the original one.
          The overall equivalence is 70\%, with a strong asymmetry across complexity levels: 92\% for low, 81\% for medium, and 47\% for high complexity features.
    \item \textit{RQ2. What is the cost involved in this migration task?}
          We collected the time spent in each migration, and the number of input and output tokens consumed by the agent during the process.
          The average migration time spent was 4min 54s.
          Low-level features consumed 1.47M tokens on average (\$1.66), whereas high-level ones consumed 9.09M tokens on average (\$10.28); 6x more.
\end{itemize}

This work makes three main contributions.
First, we present a case study on using an AI agent for legacy system modernization in a real industrial setting.
Second, we propose an evaluation methodology to objectively measure both agent \textit{effectiveness}---by cataloging persistence and functional instructions in the legacy system, and mapping them to its migrated version---and agent \textit{efficiency}---by measuring time, cost, and token consumption.
Third, we reveal practical scenarios and circumstances in which this migration strategy is more effective, as well as discuss the limitations and challenges of using AI agents for this task.

The remainder of this paper is organized as follows.
Section~\ref{sec:study-design} describes the study design, including the legacy system, the selected agent, the migrated features, and the execution workflow.
Section~\ref{sec:research-questions} presents the research questions and the evaluation metrics.
Section~\ref{sec:results} reports the main results.
Section~\ref{sec:discussion} discusses the implications of these findings.
Sections~\ref{sec:related-work} and~\ref{sec:threats} review related work and examine threats to validity.
Finally, Section~\ref{sec:conclusion} concludes the paper and points to future directions.

\section{Study Design}
\label{sec:study-design}

\subsection{Our Legacy System}

The legacy system selected for migration is a corporate Enterprise Resource Planning (ERP) used for managing shopping malls, in production for over two decades.
The software, currently implemented in VB6, is organized around a plugin architecture.
The core module gathers the features considered essential for running the business, such as users, billing and expenses management;
it currently contains 819K Lines of Code (LOC).
Secondary modules extend the ERP's core and provide specific features.
These components are implemented and provided separately, being plugged into the core according to each client request.
In this case study, two secondary modules were selected for migration:

\begin{itemize}
    \item \textsc{Measurements} is responsible for measuring the consumption of basic resources---e.g.,~water and energy---and for calculating the associated cost to each store in the mall.
          It is currently in use by 219 clients and contains around 299K LOC.
    \item \textsc{Purchases} is responsible for managing the acquisition of goods and services for the mall.
          The module's features include suppliers management, prices negotiation, purchase orders with hierarchical approval, and cost allocation.
          It is currently being used by 61 clients and contains around 224K LOC.
\end{itemize}

\begin{figure}[pos=h]
    \centering
    \includegraphics[width=\columnwidth]{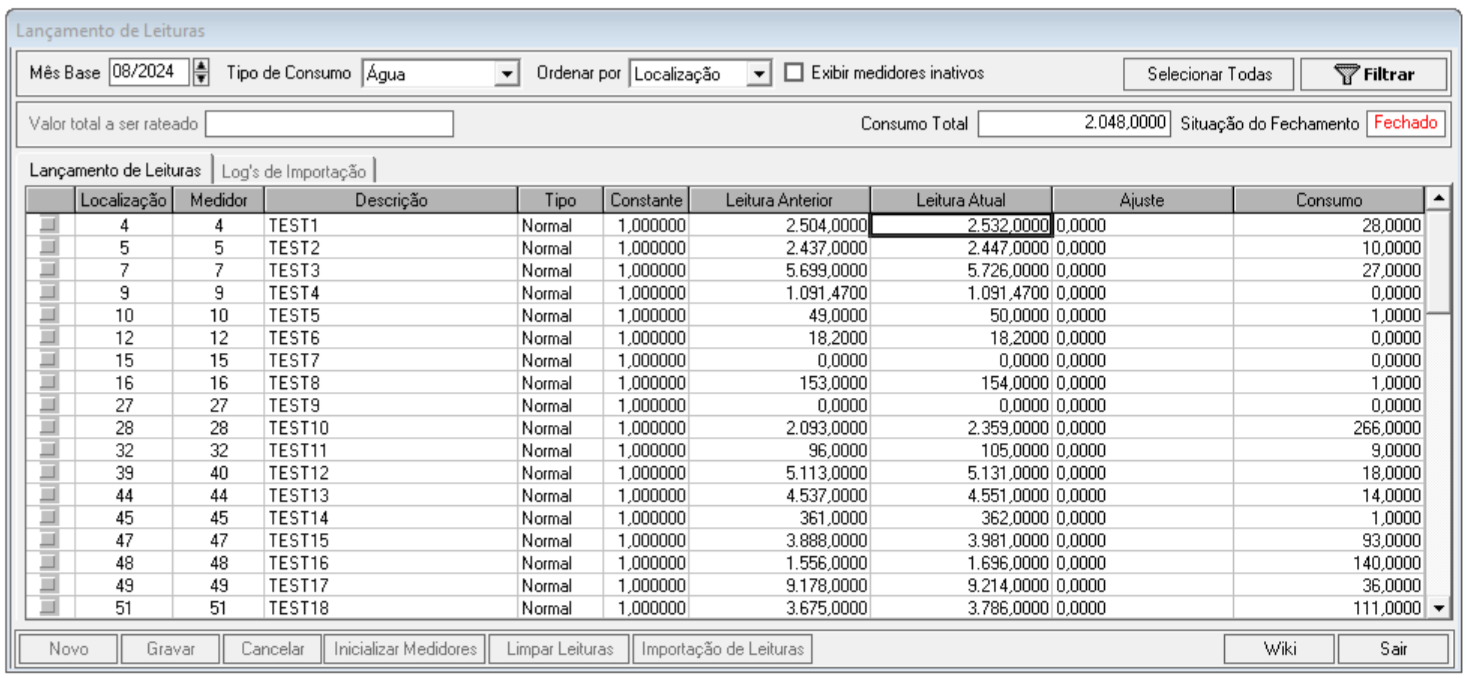}
    \caption{\textsc{Measurements} module from the Legacy System.}
    \label{fig:print-sistema}
\end{figure}

Despite being non-core modules, both components are actively and continuously used by the clients of the owning company, as these features are used in daily operations.
They cover different domains; \textsc{Measurements} is focused on numerical calculations, while \textsc{Purchases} heavily depends on transactions and approval workflows.
On the technical side, both components adopt typical structure and architectural patterns for VB6, with business logic heavily coupled to forms (\texttt{.frm}) and domain classes (\texttt{.cls}).
Altogether, these modules explore the agent's capacity under different conditions, thus allowing us to better identify its strengths and limitations.
Figure \ref{fig:print-sistema} depicts the interface of the \textsc{Measurements} module, currently showing the readings entry screen.

\subsection{The Target Migration Platform}
We defined the .NET 10 platform with C\# as our target for migration.
We planned a layered architecture design for both modules, with Controllers, Services and Repositories, and data access via Dapper to the SQL Server database.
This architecture was defined based on the previous experience of the ERP's owning company, which has been adopting this technology stack for new projects.
Finally, we deliberately maintained the same database for the migrated system to avoid further issues related to data migration.

\subsection{The Selected Agent}

We conducted the migration sessions using Claude Code (version 2.1.96), provided as a command-line tool by Anthropic.
The whole experiment was conducted with the agent operating on the Opus 4.6 1M model.
We opted for this model due to three main characteristics:
(a) a long context window capacity (up to 1 million tokens), allowing the agent to analyze the legacy codebase, as well as the target design and architectural instructions~\citep{anthropic2026opus46};
(b) direct access to the file system and terminal commands that enable the agent to create, edit, compile and execute files directly~\citep{anthropic2026claudecode};
and (c) its well-known reasoning capacity on different programming languages~\citep{anthropic2026opus46}, making it suitable for this task.

\subsection{The Selected Features}
\label{sec:selecao-funcionalidades}

We selected 12 features for this case study; six from each module.
Table \ref{tab:features} details the features selected for migration.
Besides the basic information about each feature, such as its name and description, we also report basic source code metrics used to measure the complexity of each feature;
the explanation about how we defined these complexity levels is given later in this section.
The selected features cover a wide range of complexity levels, with six features classified as low (L), four as medium (M), and two as high (H).
Finally, the last three columns describe the number of business rules and database operations implemented for each feature in the legacy system.
We rely on this information to verify the equivalence of the migrated features;
they are also described in detail next.

\begin{table*}[!htb]
    \centering
    \caption{Detailed information about the 12 features selected for migration.}
    \label{tab:features}
    \small
    \begin{tabularx}{\textwidth}{clX|rrrc|rrr}
        \toprule
        \multirow{2}{*}{\textbf{ID}}             & \multirow{2}{*}{\textbf{Feature}}         & \multirow{2}{*}{\textbf{Description}}                                        & \multicolumn{4}{c|}{\textbf{Complexity}}
                                                 & \multicolumn{3}{c}{\textbf{Instructions}}                                                                                                                                                                                                                                \\
        \cmidrule(lr){4-10}
                                                 &                                           &                                                                              & \textbf{LOC}                             & \textbf{Meth.} & \textbf{Dep.} & \textbf{Lvl} & \textbf{Rules} & \textbf{DB Ops.} & \textbf{Total} \\
        \midrule
        \multicolumn{2}{l}{\textsc{Measurement}} &                                           &                                                                              &                                          &                &               &              &                &                                   \\
        $F_{1}$                                  & Consumption Types                         & Management of consumption types (Water, Energy, Gas).                        & 1,155                                    & 38             & 1             & L            & 9              & 6                & 15             \\
        $F_{2}$                                  & Contracted Demands                        & Management of contracted demands per meter.                                  & 804                                      & 33             & 2             & L            & 6              & 6                & 12             \\
        $F_{3}$                                  & Consumption Tariffs                       & Management of tariffs per consumption type.                                  & 1,255                                    & 44             & 5             & L            & 7              & 8                & 15             \\
        $F_{4}$                                  & Meter Management                          & Meter management with hierarchy, allocation per store and cost center.       & 3,070                                    & 66             & 5             & M            & 17             & 22               & 39             \\
        $F_{5}$                                  & Mobile Devices                            & Management of mobile devices and reading synchronization.                    & 2,793                                    & 67             & 1             & M            & 12             & 17               & 29             \\
        $F_{6}$                                  & Consumption Calc.                         & Reading entry with consumption calculation and import.                       & 6,560                                    & 49             & 10            & H            & 26             & 32               & 58             \\
        \midrule
        \multicolumn{2}{l}{\textsc{Purchases}}   &                                           &                                                                              &                                          &                &               &              &                &                                   \\
        $F_{7}$                                  & Unit of Measure                           & Management of units of measure applicable to purchase items.                 & 722                                      & 18             & 1             & L            & 6              & 5                & 11             \\
        $F_{8}$                                  & Color                                     & Management of colors applicable to purchase items.                           & 842                                      & 20             & 1             & L            & 6              & 5                & 11             \\
        $F_{9}$                                  & Brand                                     & Management of brands associated with purchase items.                         & 926                                      & 20             & 1             & L            & 6              & 5                & 11             \\
        $F_{10}$                                 & Proposal Negotiation                      & Proposal negotiation rounds with unit price recalculation.                   & 4,158                                    & 63             & 14            & M            & 12             & 15               & 27             \\
        $F_{11}$                                 & Payment Method                            & Management of payment methods with terms and installment composition.        & 1,789                                    & 44             & 2             & M            & 20             & 11               & 31             \\
        $F_{12}$                                 & Purchase Request                          & Request workflow with items, hierarchical approval levels and apportionment. & 11,896                                   & 73             & 14            & H            & 33             & 39               & 72             \\
        \midrule
        \multicolumn{3}{l|}{\textbf{Total}}      & \textbf{35,970}                           & \textbf{535}                                                                 & \textbf{56}                              & ---            & \textbf{160}  & \textbf{171} & \textbf{331}                                       \\
        \bottomrule
    \end{tabularx}
\end{table*}

\noindentparagraph{Defining Complexity Levels.}
We leveraged three complexity levels---Low, Medium, and High---and associated them with each feature.
The idea is to expose the agent to different code structures, thus analyzing it under varied circumstances.
We rely on three source code metrics to determine the complexity level of each feature: lines of code, number of methods, and number of class dependencies.
Table \ref{tab:niveis-complexidade} presents the intervals for each of these metrics.

\begin{table}[pos=h]
    \centering
    \caption{Criteria for categorizing complexity levels.}
    \label{tab:niveis-complexidade}
    \small
    \begin{tabular}{l c c c}
        \toprule
        \textbf{Criterion} & \textbf{Low} & \textbf{Medium} & \textbf{High} \\
        \midrule
        Lines of Code      & Up to 1500   & 1501--5000      & $>$ 5000      \\
        Methods            & Up to 40     & 41--70          & $>$ 70        \\
        Class Dependencies & Up to 5      & 6--9            & $\geq$ 10     \\
        \bottomrule
    \end{tabular}
\end{table}

If a feature meets at least two of the three criteria for a given level, it is assigned to that level.
As we can see in Table \ref{tab:features}, low-level features mostly include simple CRUD operations, with low class dependencies;
the medium ones involve manipulating data belonging to central classes and their related records;
and the high ones deal with difficult business rules, such as complex calculations, batch processing, sophisticated workflows.

\noindentparagraph{Collecting Rules and Operations.}
The first author---who is also the software engineer responsible for the maintenance of the legacy system---performed a manual inspection over the legacy source code of each feature, and counted the number of instructions that fit into one of the following categories:

\begin{itemize}
    \item \textbf{Rules}, which include any validation, calculation, or specific restriction implemented in the code, such as business rules, data integrity checks, or domain-specific logic.
    \item \textbf{Database Operations}, instructions that interact with the system's database, such as SQL queries (SELECT, INSERT, UPDATE, DELETE) or any data manipulation that affects the persistence layer.
\end{itemize}

By counting these items in the legacy code, we can establish a baseline for comparing both legacy and migrated versions.
To illustrate this classification, Listing \ref{lst:legacy-example} presents a simplified snippet from the \textit{Consumption Types} ($F_{1}$) feature (legacy version).
Lines 2--7 map to one validation \textit{Rule} that checks if a code already exists before creation, while lines 10--13 map to one \textit{Database Operation}.

\begin{figure}[pos=h]
    \caption{Example of Rule and Database Operation categorized in Consumption Types feature ($F_{1}$).}
    \label{lst:legacy-example}
    \begin{minted}[
        % fontsize=\footnotesize,
        linenos,
        breakafter=(,
        highlightlines={2-7,10-13}
    ]{vbnet}
        ' [Rule] Business rule: Data integrity check
        If bActionFlag = F_NEW Then
            If objTiposConsumo.VerificaSeExisteCodigo(txtCodTipoConsumo.Text) Then
                MsgBox "Codigo informado ja existe.", INCONSICON, INCONS
                Exit Sub
            End If
        End If

        ' [Database Operation] Interacting with the persistence layer
        Dim Sql As String

        Sql = "SELECT NomeConsumo FROM ME_TiposConsumo " & _
            "WHERE CodTipoConsumo = '" & txtCodTipoConsumo.Text & "'"

        SelecionaTipoConsumo = ObjSeleciona.SelCodigo( _
            Sql, "Nome", "Codigo", 1)
\end{minted}
\end{figure}

As shown in Table \ref{tab:features}, we mapped 160 rules and 171 database operations across the 12 features, totaling 331 instructions.
Low-complexity features presented 11--15 instructions, while medium and high-complexity features ranged from 27--39 and 58--72 instructions, respectively.

\subsection{Migration Execution}

\begin{figure*}[pos=ht]
    \centering
    \includegraphics[width=0.65\textwidth]{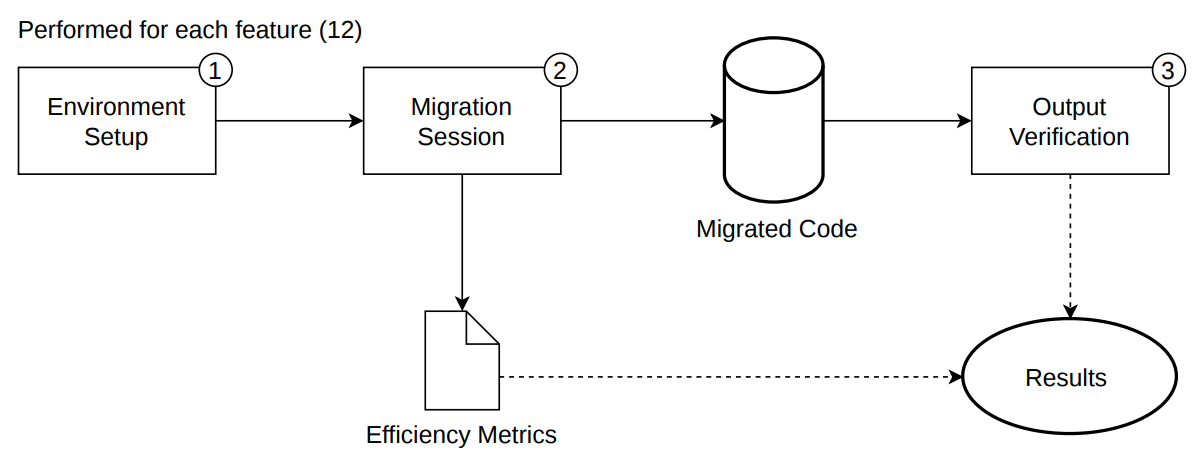}
    \caption{The methodology pipeline adopted during the migration process.}
    \label{fig:migration-execution}
\end{figure*}

Figure~\ref{fig:migration-execution} depicts the methodology adopted to perform the migration process.
We divided this process into three major steps, which are described below.

\noindentparagraph{{\small \bcircle{1}} Environment Setup.}
We set up an isolated environment for each feature, so migrations can be executed in complete independence from each other.
Each working directory is configured with the following structure\footnote{Each feature has its own set of three \texttt{CLAUDE.md} files; we provide the set of one feature as an example at \url{https://zenodo.org/records/21983175}.}:
(a) a \texttt{CLAUDE.md} file in directory root, containing general instructions for the agent, such as the migration's goal, its functional scope, migration methodology and coding conventions;
(b) a \texttt{-legacy-version} subdirectory, containing the original source code files written in VB6 (\texttt{.cls}, \texttt{.frm}), flagged as read-only so that the agent can read but not edit them, and accompanied by its own \texttt{CLAUDE.md} that describes the legacy source (its organization, the artifacts to be migrated, and the legacy coding patterns) as a reference;
and (c) a \texttt{-migration-version} subdirectory, containing the template of the target project in C\# .NET 10, with its own \texttt{CLAUDE.md} detailing specific conventions for its structure, i.e.,~the layered architecture (\texttt{Data}, \texttt{Endpoints}, \texttt{Models}, \texttt{Services}) expected for the migrated code.

\noindentparagraph{{\small \bcircle{2}} Migration Session.}
We followed a single-pass generation strategy, where the agent executes the complete migration at once, without human intervention or additional refinement after completion.
Specifically, we requested Claude to migrate the feature contained in the working directory through a single, standardized prompt.
Figure \ref{fig:claude-prompt} presents the prompt used to direct the agent during the migration sessions.

\begin{figure}[pos=h]
    \begin{promptbox}{Claude Prompt}
        \begin{minted}[
            frame=none,
            breaksymbolleft=,
            breaksymbolright=
        ]{markdown}
            Perform the migration of the functionality contained in the current directory.
        
            Structure:
            - `CLAUDE.md` (directory root) --- contains objective, scope, methodology, metrics, and conventions. Read it entirely before starting.
            - `<MODULE>-legacy-version/` --- original VB6 code (read-only, use as reference). Has its own `CLAUDE.md` describing the legacy source.
            - `<MODULE>-migration-version/` --- C# .NET project where the migration must be implemented. Has its own `CLAUDE.md` with technical details.
        \end{minted}
    \end{promptbox}
    \caption{Prompt passed to Claude Code to perform each feature migration.}
    \label{fig:claude-prompt}
\end{figure}

Note that this strategy does not prevent the agent from performing internal refinements during the session, e.g.,~reloading files, rewriting code, and adjusting the implementation as many times as needed;
instead, it does not consider human feedback between iterations~\citep{desiano2025}.
In practice, the first author played a supervisor role during the migration sessions.




\noindentparagraph{{\small \bcircle{3}} Output Verification.}
After the session ended, the first author manually inspected the generated artifacts.
Specifically, he detected the rules-based instructions in the migrated code, and matched them with the ones cataloged in the legacy code.
He also executed both versions of the system, and compared the database state produced by each one.
Finally, he collected the efficiency metrics consumed during the session, including the iteration time and number of tokens spent.

\section{Research Questions}
\label{sec:research-questions}

The goal of this work is to, from the software engineering perspective, evaluate how well coding agents can migrate legacy software to newer platforms.
Inspired by Goal-Question-Metric (GQM) approach~\citep{basili1994}, we defined two major research questions to guide our evaluation.

\subsection{RQ1. To what extent is the migrated feature equivalent to the original one?}
\label{sec:rq1}

In this question, we \textit{measure the \underline{agent's effectiveness} in producing a migrated feature} that is equivalent to the original one; i.e.,~to what extent the migrated code preserves the functional behavior of the legacy code.
In our view, a migrated feature is considered equivalent if it (a) applies the same set of business rules and validations during its execution; and (b) produces the same persistence state on the system database afterwards.
Therefore, we defined two specific questions:

\begin{description}
    \item [\textbf{RQ1.1.}]\enskip \textit{What is the \underline{persistence} equivalence between the migrated and original features?}
    \item [\textbf{RQ1.2.}]\enskip \textit{What is the \underline{functional} equivalence between the migrated and original features?}
\end{description}

\subsection{RQ2. What is the cost involved in this migration task?}

We aim to \textit{quantify the \underline{overall costs} involved in the migration process} so we can better understand in which scenarios the use of coding agents is also cost-effective.
Specifically, we unfolded this question into three, one for each cost dimension:

\begin{description}
    \item [\textbf{RQ2.1.}]\enskip \textit{What is the total \underline{time} spent by the agent to perform migration?}
    \item [\textbf{RQ2.2.}]\enskip \textit{What is the \underline{cost} consumed by the agent to perform the migration?}
    \item [\textbf{RQ2.3.}]\enskip \textit{How many \underline{tokens} are consumed by the agent to execute the migration?}
\end{description}

\subsection{Evaluation Metrics}
\label{sec:metricas}

Table~\ref{tab:metricas-gqm} presents the evaluation metrics we adopted to answer the research questions.
To answer RQ1, we defined two metrics to measure the equivalence level between the migrated and original features, one for the persistence state and another for the functional behavior.
As for RQ2, we defined three metrics responsible for quantifying the time, cost and computational effort involved in the migration process.
We detailed the definition and calculation of each metric in the following subsections.

\begin{table}[pos=ht]
    \centering
    \caption{Metrics to answer the research questions.}
    \label{tab:metricas-gqm}
    \small
    \begin{tabular}{lll}
        \toprule
        \textbf{RQ}          & \textbf{Metric}    & \textbf{Source}     \\ \midrule
        \multirow{2}{*}{RQ1} & Persistence Parity & Database            \\
                             & Functional Parity  & Source Code         \\ \midrule
        \multirow{3}{*}{RQ2} & Iteration Time     & Tool's Telemetry    \\
                             & Cost               & Model Specification \\
                             & Tokens             & Tool's Telemetry    \\
        \bottomrule
    \end{tabular}
\end{table}

\noindentparagraph{Equivalence Metrics.}
We designed two equivalence-based metrics---one based on \textit{Rules} and another based on \textit{Database Operations}---to measure how well the migrated feature preserves the behavior of the legacy code.

\begin{itemize}
    \item \textbf{Persistence Parity (RQ1.1)}: we executed the migrated and original features under the same usage scenario, and monitored the resulting state on the system database.
          We then compared each record in both versions;
          one record is considered equivalent if the same data is persisted in both versions, i.e.,~if the same values are stored in the same tables and fields.
    \item \textbf{Functional Parity (RQ1.2)}: for this metric, we analyzed the execution flows of the migrated and original features for each business rule identified in the legacy code.
          One rule is equivalent if the migrated feature behaves exactly as the original one, i.e.,~both reject the same invalid inputs for the same reasons, both adopt the same approval workflow, and both execute the same calculations for the same input data.
\end{itemize}

After counting the number of equivalent instructions for each dimension, we compute the equivalence percentage for each feature. This ratio, expressed in Equation~\ref{eq:equivalencia}, is calculated separately for persistence and functional parity.

\begin{equation}
    \label{eq:equivalencia}
    \text{\% equivalence} = \frac{\text{\# equivalent instructions}}{\text{\# total instructions}} \times 100
\end{equation}

\begin{table*}[!tb]
    \centering
    \caption{Migration results for each feature, including both equivalence and cost-based metrics.}
    \label{tab:resultados-individuais}
    \small
    \begin{tabular}{clc|ccc|crrrr}
        \toprule
        \multirow{2}{*}{\textbf{ID}}               & \multirow{2}{*}{\textbf{Feature}} & \multirow{2}{*}{\textbf{Lvl.}} & \multicolumn{3}{c|}{\textbf{Equivalence}} & \multirow{2}{*}{\textbf{Time}} & \multicolumn{3}{c}{\textbf{Tokens}} & \multirow{2}{*}{\textbf{Cost}}                                                               \\
        \cmidrule(lr){4-6} \cmidrule(lr){8-10}
                                                   &                                   &                                & \textbf{Persist.}                         & \textbf{Funct.}                & \textbf{Total}                      &                                & \textbf{In}    & \textbf{Out}   & \textbf{Total}  &         \\
        \midrule
        $F_{1}$                                    & Consumption Types                 & L                              & 100\%                                     & 78\%                           & 87\%                                & 3min 20s                       & 1.33M          & 7.7K           & 1.34M           & \$1.69  \\
        $F_{2}$                                    & Contracted Demands                & L                              & 100\%                                     & 100\%                          & 100\%                               & 2min 14s                       & 1.14M          & 6.2K           & 1.15M           & \$1.15  \\
        $F_{3}$                                    & Consumption Tariffs               & L                              & 100\%                                     & 86\%                           & 93\%                                & 3min 25s                       & 2.41M          & 12.9K          & 2.42M           & \$2.32  \\
        $F_{7}$                                    & Unit of Measure                   & L                              & 100\%                                     & 83\%                           & 91\%                                & 2min 56s                       & 1.20M          & 7.3K           & 1.21M           & \$1.61  \\
        $F_{8}$                                    & Color                             & L                              & 100\%                                     & 83\%                           & 91\%                                & 2min 28s                       & 1.33M          & 9.4K           & 1.34M           & \$1.54  \\
        $F_{9}$                                    & Brand                             & L                              & 100\%                                     & 67\%                           & 82\%                                & 2min 22s                       & 1.33M          & 8.2K           & 1.33M           & \$1.63  \\
        \midrule
        $F_{4}$                                    & Meter Management                  & M                              & 82\%                                      & 94\%                           & 87\%                                & 5min 19s                       & 7.20M          & 23.9K          & 7.22M           & \$7.53  \\
        $F_{5}$                                    & Mobile Devices                    & M                              & 82\%                                      & 50\%                           & 69\%                                & 9min 36s                       & 5.45M          & 26.3K          & 5.48M           & \$5.22  \\
        $F_{10}$                                   & Proposal Negotiation              & M                              & 60\%                                      & 92\%                           & 74\%                                & 8min 37s                       & 3.90M          & 20.0K          & 3.92M           & \$4.24  \\
        $F_{11}$                                   & Payment Method                    & M                              & 82\%                                      & 95\%                           & 90\%                                & 3min 16s                       & 2.17M          & 12.5K          & 2.18M           & \$2.91  \\
        \midrule
        $F_{6}$                                    & Consumption Calc.                 & H                              & 50\%                                      & 50\%                           & 50\%                                & 10min 1s                       & 12.29M         & 37.5K          & 12.32M          & \$11.60 \\
        $F_{12}$                                   & Purchase Request                  & H                              & 51\%                                      & 36\%                           & 44\%                                & 5min 15s                       & 5.84M          & 23.1K          & 5.86M           & \$8.85  \\
        \midrule
        \multicolumn{2}{l}{\textbf{Total Average}} &                                   & \textbf{84\%}                  & \textbf{76\%}                             & \textbf{80\%}                  & \textbf{4min 54s}                   & \textbf{3.8M}                  & \textbf{16.2K} & \textbf{3.81M} & \textbf{\$4.19}           \\
        \bottomrule
    \end{tabular}
\end{table*}

\noindentparagraph{Efficiency Metrics.}
We rely on the telemetry data provided by Claude Code \cite{anthropic2026monitoring} to automatically collect the efficiency metrics after each migration session.

\begin{itemize}
    \item \textbf{Iteration Time (RQ2.1)}: we collected the session time presented by Claude Code's interface right after the session ends.
    \item \textbf{Cost (RQ2.2)}: this metric is calculated based on the pricing table published by Anthropic on April 2026, when the experiment was conducted.
          For the Claude Opus 4.6 model, the costs are \$5{,}00/1M tokens for new input, \$6{,}25/1M tokens for cache write (TTL of 5 minutes), \$0{,}50/1M tokens for cache read and \$25{,}00/1M tokens for output.
          We apply this pricing to the number of tokens extracted from each session---described next---to calculate the overall cost for each migration.
    \item \textbf{Tokens (RQ2.3)}: we extracted the number of tokens available in each session's JSONL file, which has the detailed usage data.
          We sum up the tokens of three fields---\texttt{new input}, \texttt{cache write}, and \texttt{cache read}---to get the total input tokens, and the output tokens from its corresponding field.
\end{itemize}

\section{Results}
\label{sec:results}

Table \ref{tab:resultados-individuais} consolidates the results of all 12 migration sessions by equivalence and cost-based metrics.
The agent took 59 minutes and 45.8 million tokens to perform the migrations, at a total cost of \$50.29.
Overall, the agent scored 70\% equivalence across the 331 instructions evaluated.
On average, the agent returned 3.81M tokens with an average cost of \$4.19 per migration.
We break down these results by each RQ in the following sections.

\subsection{RQ1. Equivalence}

As previously stated in Section~\ref{sec:rq1}, we analyzed the equivalence between the migrated and original features by comparing the persistence (RQ1.1) and functional behavior (RQ1.2) of both versions.
In total, we analyzed 331 instructions; 171 are database operations and 160 are business rules.
Table~\ref{tab:equivalencia-nivel} presents the results by complexity level.

\begin{table}[pos=h]
    \centering
    \caption{Equivalence by Complexity Level}
    \label{tab:equivalencia-nivel}
    \small
    \begin{tabular}{lccccc}
        \toprule
        \textbf{C. Level}                     & \multicolumn{2}{c}{\textbf{\# Instructions}} & \multicolumn{2}{c}{\textbf{\% Equivalence}} & \textbf{Total}                                      \\
        \cmidrule(lr){2-3} \cmidrule(lr){4-5} & \textbf{Rules}                               & \textbf{DB Ops.}                            & \textbf{Persist.} & \textbf{Funct.} &               \\
        \midrule
        Low                                   & 40                                           & 35                                          & 100\%             & 83\%            & 92\%          \\
        Medium                                & 61                                           & 65                                          & 77\%              & 85\%            & 81\%          \\
        High                                  & 59                                           & 71                                          & 51\%              & 42\%            & 47\%          \\
        \midrule
        \textbf{Total}                        & \textbf{160}                                 & \textbf{171}                                & \textbf{71\%}     & \textbf{69\%}   & \textbf{70\%} \\
        \bottomrule
    \end{tabular}
\end{table}

\noindentparagraph{RQ1.1. What is the persistence equivalence between the migrated and original features?}
For low-level features, the agent achieved 100\% parity effectiveness, with all 35 operations correctly reproduced.
For medium-level ones, the persistence rate dropped to 77\%, as 15 out of 65 operations were not correctly migrated, mostly involving multi-table records, cascading updates, and complex filters.
The parity for high-level features stayed at 51\%, with 35 out of 71 operations being incorrectly migrated.
At this particular level, we observe a qualitative shift in the type of errors performed by the agent;
complete persistence modules are left without implementation, such as the entire reading-approval workflow of the \textit{Consumption Calculation} feature ($F_{6}$)---querying pending readings, approving them, and canceling previous approvals---which was absent altogether, rather than implemented incorrectly.

\noindentparagraph{RQ1.2. What is the functional equivalence between the migrated and original features?}
The agent achieved 83\% parity for low-level features, with 33 out of 40 rules correctly implemented;
mostly missing rules correspond to form validations dispersed in interface events of the legacy code.
Interestingly, 85\% of medium-level features were correctly migrated, i.e., 52 out of 61.
Three out of the four medium-level features achieved very high scores, i.e.,~$F_{4}$ with 94\%, $F_{6}$ with 92\% and $F_{7}$ with 95\% (Table \ref{tab:resultados-individuais}).
The rules of these features are concentrated in well-structured \texttt{.cls} classes, facilitating the agent's contextual understanding.
On the other hand, feature $F_{5}$ has rules distributed among multiple class files, hence scoring 50\% functional parity.
Finally, the agent's performance dropped to 42\% on high-level features, i.e.,~the agent was able to migrate only 25 out of the 59 rules.
Mostly missing rules involve sophisticated workflows and cross-module validations.

\noindentparagraph{Sanity Check.}
Since functional parity relies on human judgment, a second reviewer---a mid-level software engineer who did not participate in the migration sessions---independently evaluated the same 160 business rules migrated by the agent.
This second review reached almost the same aggregate result, 110 out of 160 rules classified as equivalent, i.e.,~69\% functional equivalence.
When analyzing individually, both reviewers agreed on 149 out of the 160 rules (93\%).
We also computed the Cohen's Kappa coefficient to measure inter-rater reliability, which resulted in 0.84, indicating a strong agreement between the two reviewers.

The 11 divergences concentrate on partially migrated rules, where the agent produced the rule but did not wire it into the execution flow, or preserved its intent while narrowing its criteria.
For instance, the agent correctly reimplemented the IMEI uniqueness check in $F_{5}$, yet exposed it as a standalone endpoint that neither the creation nor the update flow invokes.
While the first author counted it as divergent, the reviewer counted it as equivalent.

\begin{rqsummary}{Finding \#1}
    The agent achieved an overall equivalence of 70\%.
    This performance varies according to the complexity level, 92\% for low-level features, 81\% for medium-level ones, and 47\% for high-level features.
\end{rqsummary}



\subsection{{Understanding Failed Cases}}
The agent was not able to migrate 100 out of the 331 instructions evaluated.
In order to better understand the context of these failures, the first author manually analyzed each divergent instruction and classified them into five categories, as shown in Table~\ref{tab:divergencias}.

\begin{table}[pos=h]
    \centering
    \caption{Failure reasons and their frequency.}
    \label{tab:divergencias}
    \small
    \begin{tabular}{lr}
        \toprule
        \textbf{Issue}             & \textbf{\#}  \\
        \midrule
        Missing Adjacent Modules   & 42           \\
        Missing DB Operation       & 22           \\
        Missing Rule               & 20           \\
        Missing Module Integration & 10           \\
        Bad Implementation         & 6            \\
        \midrule
        \textbf{Total}             & \textbf{100} \\
        \bottomrule
    \end{tabular}
\end{table}

Most divergences correspond to \textit{missing adjacent modules} (42), i.e.,~ones that coexist with the core functionality in the legacy code, but the agent failed to detect and migrate them during the process.
These modules include batch data-import procedures, report generation, email notifications, etc.
Such failures happened almost exclusively in high-level features, since they are inherently more complex and contain more methods and dependencies.
In this case, the agent appears to prioritize the source code implemented in the feature's core, but does not expand this process to adjacent modules, even when they are provided as context to the agent.

\textit{Missing DB operations} (22) and \textit{missing rules} (20) are the second and third most common types of divergences, respectively.
They represent specific instructions not implemented by the agent, such as a particular validation or a specific database query.
Differently from the previous category, here the agent does cover the feature's core but overlooks individual instructions outside the main domain logic, such as validations coded throughout the form's events or checks embedded in conditional flows.

\textit{Missing module integration} (10) includes instructions that connect the feature to other modules that are outside the feature's main functions---such as sending e-mail notifications to users---and that the agent left unimplemented by concentrating only on the feature's core behavior.
Finally, the agent executed \textit{bad implementation} in just six cases.
This failure corresponds to migrations that are incorrect or incomplete, such as a validation that verifies a different condition than the original one, or a database query that updates the wrong table.
Two examples illustrate these failures.

\noindentparagraph{Mobile Devices ($F_{5}$).}
This medium-level feature achieved 69\% equivalence, with 26 out of 65 instructions not correctly migrated.
When analyzed in detail, we found that the agent correctly implemented the major execution of the feature, but failed to detect the validation rules implemented for the input form fields.
These rules were originally coded throughout different form events, outside the domain classes the agent prioritized when migrating the feature.
The agent also failed to implement a database operation that depended on checking the existence of a record before deciding between creating or updating it.
As a consequence, the migrated version duplicated the records instead of updating them.
This last issue is depicted in Figure \ref{fig:failure-example}; divergent lines are highlighted.

\begin{figure}[pos=h]
    \caption{Incorrect DB operation implemented when migrating Mobile Devices feature ($F_{5}$).}
    \label{fig:failure-example}
    \begin{minted}[
        %   fontsize=\footnotesize,
            linenos,
            highlightlines={3-4,8-15}
        ]{csharp}
public async Task<bool> UpsertDeviceSyncAsync(UpsertDeviceSyncRequest request)
{
    // Missing the existence check the legacy performs before
    // deciding between INSERT and UPDATE
    var nextId = await db.MobileDeviceSyncs
        .MaxAsync(x => (int?)x.IDSincronizacao) ?? 0;

    db.MobileDeviceSyncs.Add(new MobileDeviceSync
    {
        IDSincronizacao = nextId + 1,
        IDFilial = (short)request.IDFilial,
        IDDispositivo = request.IDDispositivo,
        TipoSincronizacao = request.TipoSincronizacao,
        DtUltSincronizacao = DateTime.Now
    });

    await db.SaveChangesAsync();
    return true;
}
    \end{minted}
\end{figure}

\noindentparagraph{Purchase Request ($F_{12}$).}
This high-level feature achieved 44\% equivalence, with 34 out of 71 instructions not correctly migrated.
This is the largest feature of the experiment---around 12K LOC, according to Table~\ref{tab:features}---and clearly illustrates the agent's limitation at detecting larger scopes.
In this case, the agent correctly migrated the feature's main execution flow, but left out alternative ones such as calling other modules and triggering automatic notifications.
The agent also simplified a multi-step process by omitting intermediate instructions, ultimately breaking the original behavior.

Figure \ref{fig:failure-example2} presents an example of these failures. As we can see, the legacy save routine performs several steps: besides persisting the request, it also contacts other users when the request is approved by its authority.
This step is not merely informative---it leaves the request in a pending state whose progression depends on the approval of the users who receive the e-mail, so it can only move forward once they act on it.
The migrated version maintained only the main workflow: it inserts the request and its items, always in the \texttt{Open} state, and never sends the notification.
The request is therefore created regardless of this authority approval.

\begin{figure*}[ht]
    \centering
    \caption{Incorrect implementation of a multi-step process when migrating Purchase Request feature ($F_{12}$).}
    \label{fig:failure-example2}
    \begin{subfigure}[t]{0.6\linewidth}
        \caption{Legacy code (VB6)}
        \label{lst:failure-example2-legacy}
        \begin{minted}[
            %   fontsize=\footnotesize,
                linenos,
                highlightlines={5,8-16}
            ]{vbnet}
                With objSolicitacao
                    ' Insert request (Finalized if item approval is bypassed)
                    lngIdSolicitacao = .Incluir(gIDSistema, gIDFilial, _
                        gCategoria, gIdUsuario, ..., _
                        IIf(cfg_IgnoraAprovacaoItens, adFinalizada, adAberta), _
                        dbValorSolicitacao, ...)

                    ' When approval tiers are enabled, e-mail the approvers
                    ' (the request stays pending until they approve it)
                    If objConfiguracaoInicialGlobal.AprovacaoAlcadaSolicitacao Then
                    If objConfiguracaoInicialGlobal.EmailAlcSolicitacao Then
                        .EnvioEmail_UsuarioAprovador gIDSistema, gIDFilial, _
                            gCategoria, dbValorSolicitacao, lngIdSolicitacao, _
                            adSolicitacaoCompra, adAprovacaoItem
                    End If
                    End If

                    ' Persist the request items
                    Gravar_Dados_Planilha CLng(lngIdSolicitacao), DadosOutSolicacao
                End With
            \end{minted}
    \end{subfigure}

    \quad

    \begin{subfigure}[t]{0.6\linewidth}
        \caption{Migrated code (C\#)}
        \label{lst:failure-example2-migrated}
        \begin{minted}[
            %   fontsize=\footnotesize,
                linenos,
                highlightlines={4,15}
            ]{csharp}
                var request = new PurchaseRequest
                {
                    // ...
                    Status = PurchaseRequestStatus.Open
                };
                db.PurchaseRequests.Add(request);

                // Insert the request items
                foreach (var itemDto in dto.Items)
                    db.PurchaseRequestItems.Add(new PurchaseRequestItem { /* ... */ });

                request.RequestedAmount = totalValue;
                await db.SaveChangesAsync();
                
                // No approver e-mail, no bypass/auto-authorize handling
            \end{minted}
    \end{subfigure}
\end{figure*}

\begin{rqsummary}{Finding \#2}
    Most failures correspond to source code instructions ignored by the agent, despite being present in the legacy code provided as context.
    The agent overlooked adjacent modules (42 cases), did not consider specific DB operations (22) and missed particular rules (20).
\end{rqsummary}

\subsection{RQ2. Efficiency}

We analyzed the agent's efficiency under three dimensions: time (RQ2.1), cost (RQ2.2), and tokens consumed (RQ2.3). Table~\ref{tab:eficiencia-nivel} aggregates the results of efficiency-based metrics by complexity level.

\begin{table}[pos=h]
    \centering
    \caption{Agent efficiency by complexity level.}
    \label{tab:eficiencia-nivel}
    \small
    \begin{tabular}{l r c r r r}
        \toprule
        \multirow{2}{*}{\textbf{Level}} & \multirow{2}{*}{\textbf{Duration}} & \multirow{2}{*}{\textbf{Cost}} & \multicolumn{3}{c}{\textbf{Tokens}}                                    \\ \cmidrule(lr){4-6}
                                        &                                    &                                & \textbf{Input}                      & \textbf{Output} & \textbf{Total} \\ \midrule
        Low                             & 2min 48s                           & \$1.66                         & 1.46M                               & 8.6K            & 1.47M          \\
        Medium                          & 6min 42s                           & \$4.97                         & 4.68M                               & 20.7K           & 4.70M          \\
        High                            & 7min 38s                           & \$10.23                        & 9.06M                               & 30.3K           & 9.09M          \\
        \midrule
        \textbf{Average}                & \textbf{4min 54s}                  & \textbf{\$4.19}                & \textbf{3.80M}                      & \textbf{16.2K}  & \textbf{3.81M} \\
        \bottomrule
    \end{tabular}
\end{table}

\noindentparagraph{RQ2.1. What is the total time spent by the agent to perform migration?}
The total average time spent by the agent to perform migration was 4 minutes and 54 seconds, with a clear distinction between low-level features and medium/high-level ones.
While the agent spent 2 minutes and 48 seconds on average for low-level features, it took 2.4x for medium (6min 42s) and 2.7x for high-level features (7min 38s).

\noindentparagraph{RQ2.2. What is the cost consumed by the agent to perform migration?}
The average cost was \$4.19 per migration.
When we break down by complexity level, we observe an even greater contrast when compared to migration time.
The cost increased by 2.9x for medium-level features (\$4.97) and 6.2x for high-level ones (\$10.23).
Still, the cost-benefit ratio remains favorable for using agents in this context, as it migrated 331 instructions at a total cost of \$50.29, i.e.,~\$0.15 per instruction.

\noindentparagraph{RQ2.3. How many tokens are consumed by the agent to perform migration?}
The agent consumed 3.81M tokens per migration in total.
Input tokens dominate the consumption (99.6\% of the total), as the agent rereads the legacy code at each reasoning cycle, while the generated output is comparatively compact.
This refinement helps keep the cost of the task at a reasonable level, as input tokens are cheaper than output tokens.
The same disproportionate difference persists among low and medium/high-level features.
While the average token consumption for low-level features was 1.47M, it increased by 3.2x for medium-level features (4.70M) and 6.2x for high-level ones (9.09M).

\subsection{{Cost-Benefit Analysis}}

\begin{figure}[pos=h]
    \centering
    \begin{tikzpicture}
        \begin{axis}[
                scatter,
                only marks,
                mark size=3pt,
                xlabel={Cost (USD)},
                ylabel={Equivalence (\%)},
                xmin=0, xmax=14,
                ymin=30, ymax=105,
                grid=both,
                grid style={dashed, gray!30},
                width=1\columnwidth,
                height=5.0cm,
                point meta=explicit symbolic,
                scatter/classes={
                        low={mark=*, blue!70, draw=black, line width=0.1pt},
                        medium={mark=square*, orange!80, draw=black, line width=0.1pt},
                        high={mark=triangle*, red!70, draw=black, line width=0.1pt}
                    },
                legend style={at={(0.98,0.98)}, anchor=north east, font=\scriptsize},
            ]
            \addplot[scatter, scatter src=explicit symbolic]
            coordinates {
                    (1.69,87)  [low]
                    (1.15,100) [low]
                    (2.32,93)  [low]
                    (1.61,91)  [low]
                    (1.54,91)  [low]
                    (1.63,82)  [low]
                };
            \addplot[scatter, scatter src=explicit symbolic]
            coordinates {
                    (7.53,87) [medium]
                    (5.22,69) [medium]
                    (4.24,74) [medium]
                    (2.91,90) [medium]
                };
            \addplot[scatter, scatter src=explicit symbolic]
            coordinates {
                    (11.60,50) [high]
                    (8.85,44)  [high]
                };
            \legend{Low, Medium, High}
        \end{axis}
    \end{tikzpicture}
    \caption{Cost vs. Equivalence by Complexity Level.}
    \label{fig:custo-vs-equivalencia}
\end{figure}
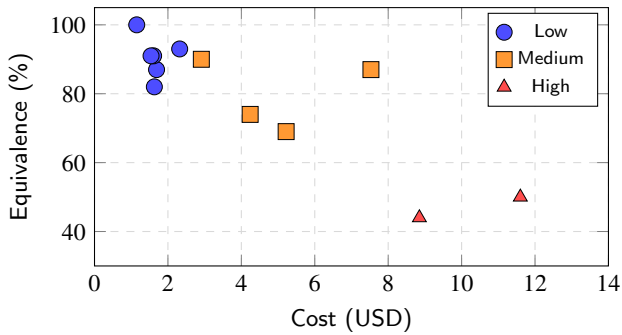

Figure \ref{fig:custo-vs-equivalencia} presents a cost-benefit analysis of the agent's efficiency, considering its equivalence and cost across the three complexity levels.
A clear trade-off emerges between feature complexity and the agent's ability to migrate them.
The agent migrated low-level features with both high equivalence (92\%, average) and low cost (\$1.66, average), whereas high-level ones achieved 47\% equivalence at a cost of \$10.23.
Put differently, the agent performed 1.98x better at a fraction of 0.16x of the cost
when compared to high-level features.
The results for medium-level features are mixed.
Two of them ($F_{4}$ and $F_{11}$) achieved high equivalence---87\% and 90\%, respectively---at a moderate cost---\$7.53 and \$2.91, respectively---while the other two ($F_{5}$ and $F_{10}$) scored below 80\% equivalence and cost above \$4.00.

\begin{rqsummary}{Finding \#3}
    The migration cost is largely determined by the feature's size and complexity on the legacy version.
    Low-level features cost \$1.66 on average, whereas medium and high-level ones cost \$4.97 and \$10.23.
\end{rqsummary}

\section{Discussion}
\label{sec:discussion}

\subsection{Agents Performance on Fragmented Code}

While we obtained functional equivalence rates above 90\% for three out of the four medium-level features---$F_{4}$ (94\%), $F_{10}$ (92\%), and $F_{11}$ (95\%)---only one of the six low-level ones achieved a similar level ($F_{2}$, with 100\%).
As a result, the average \textit{functional equivalence for medium-level features (85\%) was higher than that of low-level ones (83\%)}, which is counterintuitive given their difference in complexity and instructions.
For instance, $F_{9}$ (Brand) is a low-level feature with only 926 LOC and six rules, whereas $F_{10}$ (Proposal Negotiation) is a medium-level one with 4,158 LOC and 12 rules;
yet $F_{9}$ achieved only 67\% feature equivalence and $F_{10}$ reached 92\%.

Looking beyond the size of the features, we found that all three medium-level ones with high equivalence have their rules concentrated in well-scoped domain classes (\texttt{.cls}),
thus allowing the agent to effectively capture these rules.
On the other hand, we noted that most of the missing rules in low-level features correspond to behaviors coded outside the domain classes, such as form interface events (\texttt{.frm}) or referenced via global objects.
That is exactly the case of $F_{9}$, whose missing rule is a query filter (\texttt{IdMarca > 0}) coded in the form that loads the brand list.
This filter discards invalid records and, as it is not declared in the domain class, the agent left it out of the migrated version.
Likewise, $F_{5}$ is the only medium-level feature with low feature equivalence rate (50\%);
the missing rules were also coded directly in its form, and thus overlooked by the agent.
Therefore, fragmented code across multiple files---even when the files are completely provided---appears to be a key determinant of the agent's performance, echoing similar findings in the literature~\citep{Zhang2023b, Jimenez2024}.

\subsection{Agent Favors Explicit Rules over Implicit Ones}

We also observed that the agent's effectiveness varied depending on how the instructions were codified in legacy code.
On one hand, the agent successfully migrated \textit{explicit} instructions, those directly implemented in the legacy code as method validations, properly identified database operations, etc.
On the other, it ignored \textit{implicit} instructions, those hidden in optional query parameters, dynamic configuration values, or form-level conditions that silently change the system's behavior.

For instance, the agent failed to migrate two rules in $F_{1}$ (Consumption Types), neither declared as an explicit validation. The first is controlled by the \texttt{NaoEnergia} parameter, a boolean argument that excludes the Energy type (\texttt{E}) from the result when enabled.
The second blocks the deletion of consumption codes defined either by the system's default-values or by the \texttt{cfg\_codigotr} global variable, whose value is dynamically loaded from the database at startup.
Likewise, the agent omitted query filters with implicit thresholds in $F_{8}$ (\texttt{IdCor > 0}) and $F_{9}$ (\texttt{IdMarca > 0}). This suggests that developers should avoid delegating the inference of important restrictions to the agent, as it can easily overlook them.
Instead, such information should be stated as clear and explicit as possible.

\subsection{Migration Cost is Driven by Feature's Size}

With respect to agent efficiency, we observed an asymmetry between the cost of input and output tokens.
Despite output tokens costing 5x more (\$25.00 vs. \$5.00 per 1M tokens), the agent consumes overwhelmingly more input than output tokens.
Such discrepancy is noted across all complexity levels (see Table~\ref{tab:eficiencia-nivel});
the agent consumed on average 1.46M input tokens and 8.6K output tokens (170x) for low-level features, 4.68M and 20.7K (226x) for medium-level ones, and 9.06M and 30.3K (299x) for high-level features.

Put simply, the cost of a migration is largely affected by the volume of what it receives; not by the size of what it generates.
This behavior is clearly noted in Figure \ref{fig:custo-vs-equivalencia}, where the cost of three features stands out as outliers:
one is $F_{4}$, the medium-level feature that consumed 7.20M input tokens; the other two are $F_{6}$ and $F_{12}$, the high-level features that consumed 12.29M and 5.84M input tokens, respectively.

\section{Related Work}
\label{sec:related-work}

\subsection{Software Engineering supported by LLMs}

Since the emergence of LLMs, we have observed an increasing number of studies evaluating their application to software engineering tasks~\citep{Silva2024, Terragni2024, Sobania2023, Le2024, Improta2025, Chen2024, Caumartin2025, Guelman2026, Alves2026}.
\citet{hou2024} conducted a systematic literature review with 395 papers on LLMs applied to software engineering, mapping the state of the art between 2017 and 2023.
Code generation and comprehension tasks are the most explored, while legacy system modernization remains underrepresented.
Our work addresses this gap by empirically evaluating the modernization of enterprise systems maintained for decades on outdated languages such as VB6.

\subsection{Software Migration supported by LLMs}

Software migration is crucial for modern software development as it involves adapting the software to ensure long-term functionality and compatibility~\citep{Ziftci2025}.
Such procedure can happen in different scopes.
At a narrow level, developers can migrate a single API to a new version~\citep{Almeida2024, Islam2024, Lamothe2021, Alves2026}, whereas at a broader level they can update an entire system to a new language or platform~\citep{Macedo2025,Mendonca2024,Jimenez2024}.
In our perspective, the approach described in this paper fits in the second category as it involves migrating a whole component from one language to another.

Previous studies have also explored the use of LLMs for migrating software at this level, i.e.,~wider scope.
\citet{yang2024} and \citet{feischl2025} evaluate the ability of LLMs to translate code between languages such as Python, Java, C++ and Rust, using test suites to verify the functional equivalence.
Similar to our findings, both studies report that model performance is constrained by the level of complexity of the code being migrated.
They also observe that the generated code can differ from the original behavior in edge cases, even when it compiles and runs successfully.
While these studies evaluate the translation of code snippets in modern languages, our work focuses on migrating legacy systems, i.e.,~systems implemented in discontinued languages such as VB6.

\citet{desiano2025} analyzed how this language translation process can be improved with human-in-the-loop, showing that introducing iterative cycles significantly increases the equivalence of the migrated code.
In an opposite direction, we adopted a single-generation strategy since we are assessing agents---not LLMs---efficiency, and thus wanted to consider the agent's autonomous capabilities in the analysis.

\section{Threats to Validity}
\label{sec:threats}


\noindentparagraph{Construct Validity.}
We collected the equivalence metrics manually, which introduces the risk of subjectivity when classifying items as equivalent or divergent.
To reduce this risk, we cataloged each rule and database operation before the migration sessions and defined equivalence criteria upfront.
We also conducted a sanity check where a second reviewer independently evaluated the migration performed over all 160 business rules.
Both reviewers reached the same overall functional equivalence with 0.84 inter-rater agreement, indicating a strong inter-rater reliability (Section~\ref{sec:results}).
We extracted the cost-based metrics from Claude Code's session telemetry.

\noindentparagraph{Internal Validity.}
We categorized features as low, medium, or high complexity using three objective criteria (LOC, number of methods, and number of class dependencies).
However, some features fall near the boundaries, and a slightly different threshold could move their classification.
Requiring at least two out of three to define the complexity level mitigates this risk.
We also note that we deliberately used a single-generation strategy; our results should not be extrapolated to iterative approaches, which tend to produce higher equivalence~\citep{desiano2025}.

\noindentparagraph{External Validity.}
We conducted this study on two modules of a single VB6 ERP, migrating them to C\# .NET 10 with a single agent (Claude Code 2.1.96) running a single model (Opus 4.6 1M).
We fixed this configuration on purpose, as it was the most suitable setup for the industrial context under study (Section~\ref{sec:study-design}) and keeps the sessions comparable to each other. Even so, our findings reflect this specific configuration, source-to-target language pair, and application domain. Generalizing to other agents and models, or to other combinations such as COBOL to Java or Delphi to TypeScript, requires replication; we expect the observed trends to be more stable than the absolute equivalence, time, and cost values.

\noindentparagraph{Conclusion Validity.}
Although 12 features (six low, four medium, and two high complexity) is appropriate for a case study, having only two high-complexity features limits the robustness of our conclusions for migrations at that level.
We treat the observed trends as indicative, and expanding the sample is a natural next step.

\section{Conclusion}
\label{sec:conclusion}

Legacy system modernization is a critical yet challenging task for organizations.
We report a case study on using an AI coding agent to migrate two modules of an industrial legacy system from VB6 to C\# .NET 10.
By selecting 12 features of varying complexity, we evaluated the agent's effectiveness in reproducing the original system's behavior and its efficiency in terms of time, cost, and token consumption.

Overall, the agent achieved 70\% functional equivalence after migrating all 12 features for approximately 59 minutes, and at a cost of \$50.29.
We observed a clear trade-off when breaking down by complexity level.
While low-level features reached 92\% equivalence at an average cost of \$1.66, high-level features scored only 47\% at a cost of \$10.23.
After analyzing the failed cases, we noted that features fragmented across multiple files and implicitly coded business rules detracted the agent's performance, corroborating similar findings in the literature~\citep{Zhang2023b, Jimenez2024}.
All in all, we see promising results for using AI coding agents in legacy system modernization, especially for low and well-scoped features.

\noindentparagraph{Future Work.}
Our next steps include (a) replicating this study with an iterative generation strategy to evaluate how much the agent's performance can be improved with human-in-the-loop~\citep{desiano2025};
(b) exploring a multi-agent approach that covers not only implementation but also documentation, build steps, and testing;
(c) investigating upfront decomposition of high-complexity features into smaller units to improve agent performance; and
(d) comparing results across multiple coding agents and different source-to-target language pairs.

\noindentparagraph{Replication Package.}
The replication package for this study is available at \url{https://zenodo.org/records/21983175}.

\section*{Acknowledgments}
This work was supported by Group Software, which provided the legacy systems used as the object of study. The authors also thank João Pedro for his contribution as the second reviewer in the sanity check conducted in this study. This work was partially supported by INES.IA (National Institute of Science and Technology for Software Engineering Based on and for Artificial Intelligence) \texttt{\url{www.ines.org.br}}, CNPq grant 408817/2024-0.
It was also supported by grants from CNPq (444507/2024-8, 403304/2025-3) and by FAPEMIG (APQ-02419-23).

\bibliographystyle{cas-model2-names}
\bibliography{main}

\end{document}